\documentclass[9pt]{article}
\usepackage{spconf,amsmath,graphicx}
\usepackage{caption}
\usepackage{subcaption}
\usepackage{multirow}
\usepackage{pifont}
\newcommand{\cmark}{\ding{51}}%
\newcommand{\xmark}{\ding{55}}%
\usepackage{amsmath}

\newcommand{\comment}[1]{}

\title{Emformer: Efficient Memory Transformer Based Acoustic Model For Low Latency Streaming Speech Recognition}
\name{
\begin{tabular}{c}
Yangyang Shi, Yongqiang Wang, Chunyang Wu, Ching-Feng Yeh, Julian Chan, \\
Frank Zhang, Duc Le, Mike Seltzer
\end{tabular}
}
\address{Facebook AI}
\begin{document}
\ninept
\maketitle
\begin{abstract}
This paper proposes an \textit{e}fficient \textit{m}emory trans\textit{former} \textit{Emformer} for low latency streaming speech recognition. In Emformer, the long-range history context is distilled into an augmented memory bank to reduce self-attention's computation complexity. A cache mechanism saves the computation for the key and value in self-attention for the left context. Emformer applies a parallelized block processing in training to support low latency models. We carry out experiments on benchmark LibriSpeech data.
Under average latency of 960 ms, Emformer gets WER $2.50\%$ on test-clean and $5.62\%$ on test-other. Comparing with a strong baseline augmented memory transformer (AM-TRF), Emformer gets $4.6$ folds training speedup and $18\%$ relative real-time factor (RTF) reduction in decoding with relative WER reduction $17\%$ on test-clean and $9\%$ on test-other. For a low latency scenario with an average latency of 80 ms, Emformer achieves WER $3.01\%$ on test-clean and $7.09\%$ on test-other. Comparing with the LSTM baseline with the same latency and model size, Emformer gets relative WER reduction $9\%$ and $16\%$ on test-clean and test-other, respectively.
\end{abstract}
\vspace{-7pt}
\section{Introduction}
\label{sec:intro}
\vspace{-5pt}
Transformers~\cite{vaswani_2017} have achieved dominated performance for various tasks in natural language processing area~\cite{dai_2019,devlin_2018,Raffel_2019}. Rather than using memory state to capture long-range dependencies in recurrent neural networks, the multi-head self-attention method connects arbitrary positions in the whole sequence directly in parallel. 

Recently, transformer-based model architectures have also been successfully applied to automatic speech recognition (ASR) area across various modeling paradigms, including sequence-to-sequence~\cite{dong2018speech,karita2019comparative,sperber2018self,zhou2018syllable,chengyi_wang_2020}, neural transducer~\cite{zhang_2020,Yeh_2019,anmol_gulati_2020}, Connectionist temporal classification (CTC)~\cite{Salazar2019_ICASSP,frank_zhang_2020} and traditional hybrid~\cite{povey2018time,yongqiang_2019_icassp} systems. 

Unlike most natural language processing tasks, many ASR applications deal with streaming scenarios challenging for vanilla transformers. The streaming recognizer needs to produce output given partially available speech utterance rather than entire utterance. Several methods advance the transformer for streaming speech recognition. The work~\cite{povey2018time,zhang_2020,moritz2020streaming} proposed to constrain the attention computation with a limited length of look-ahead inputs. However, these methods have a significant delay due to the look-ahead context leaking issue where essential look-ahead context grows linearly with the number of transformer layers stacking on top of one another. A scout network is proposed in~\cite{chengyi_wang_2020} to detect the word boundary.  In scout networks, only the context information before the word boundary is used by the transformer to make predictions. However, the scout network does not address the heavy self-attention computation that grows quadratically with the left context length. A streaming transformer with augmented memory (AM-TRF) is proposed in~\cite{Chunyang_2020_interspeech} to reduce latency and the self-attention computation.

AM-TRF uses a similar block processing method as~\cite{dong2019self}. The block processing chunks the whole utterance into multiple segments. To reduce the computation in capturing the long-range left context, AM-TRF introduces a memory bank. Each vector in the memory bank is an abstract embedding from the previous one segment. The direct left context block from the current segment and look-ahead context block provides context information for current segment recognition in addition to the memory bank. However, AM-TRF has duplicated computations for the direct left context block in both training and decoding. The memory bank carries over the context information from previous segments in a similar auto-regression way as recurrent neural networks. The inherent auto-regression characteristic makes AM-TRF challenging to parallelize the block processing in training.
\begin{figure*}[htbp]
\begin{subfigure}{.5\linewidth}
\centering
\includegraphics[scale=0.36]{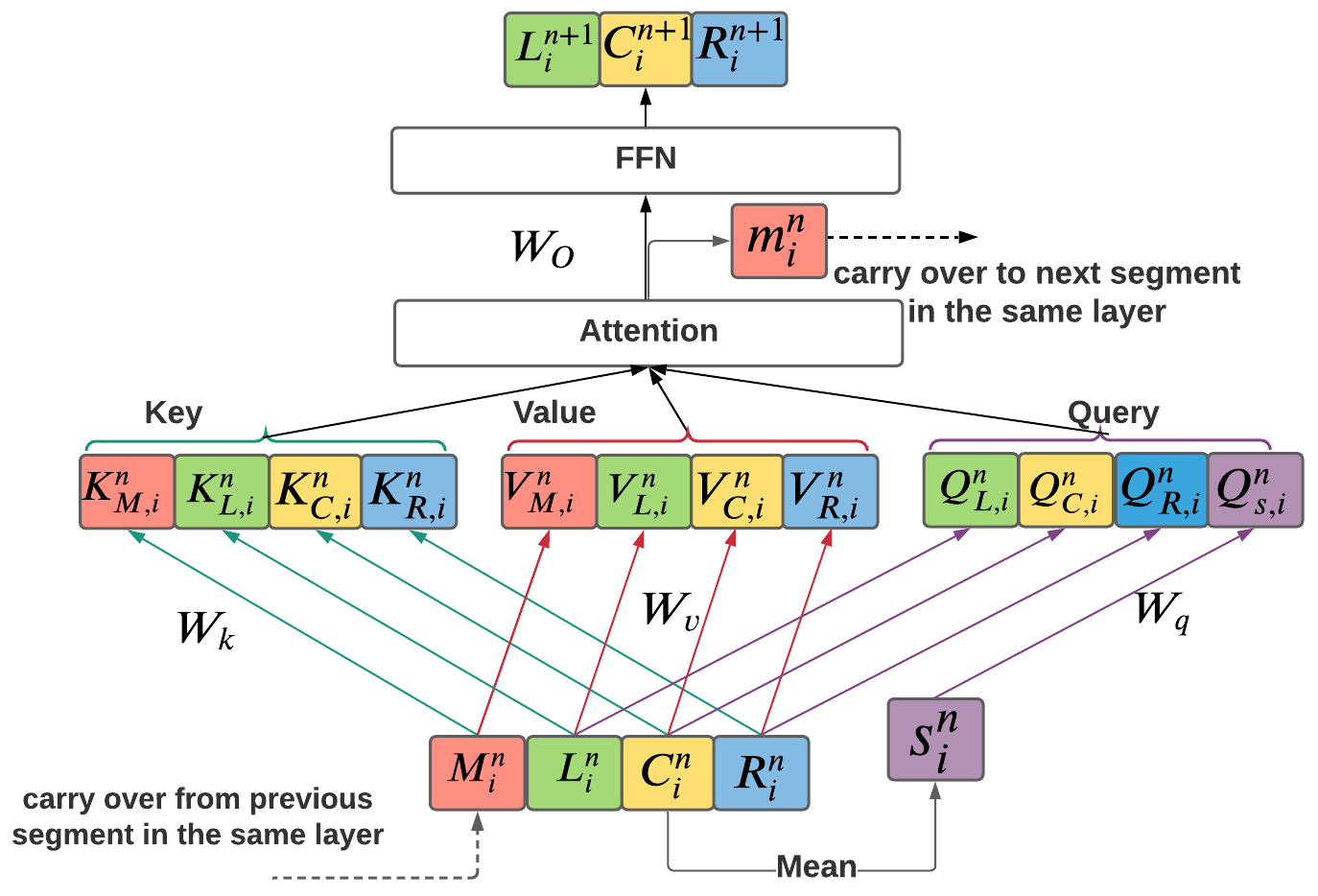}
\caption{AM-TRF}
\label{fig:amtrf}
\end{subfigure}%
\begin{subfigure}{.5\linewidth}
\centering
\includegraphics[scale=0.36]{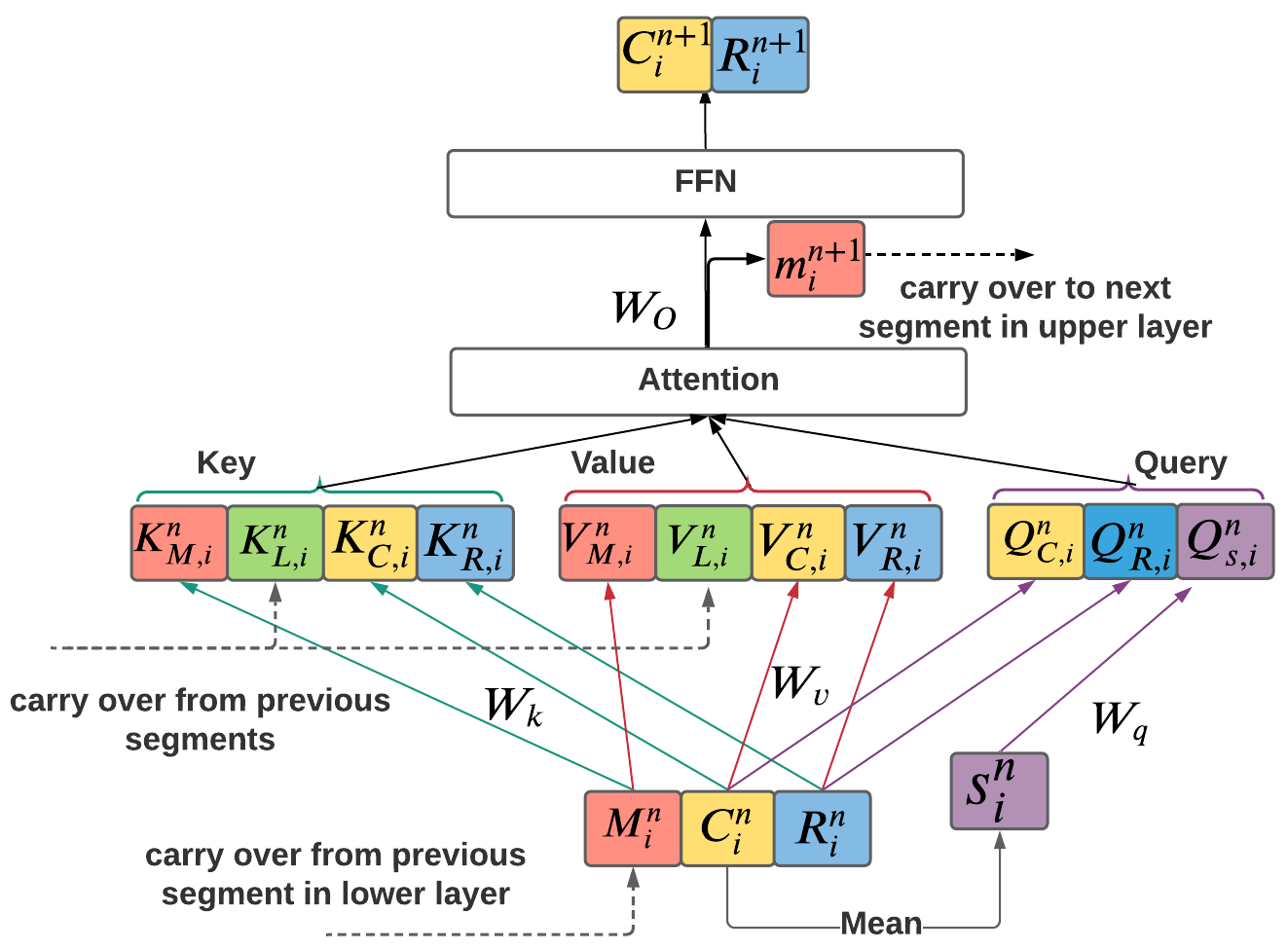}
\caption{Emformer}
\label{fig:memformer}
\end{subfigure}
\caption{Comparison of AM-TRF with Emformer}
\vspace{-15pt}
\end{figure*}

In this paper, we propose the Emformer that improves the AM-TRF from the following aspects. First, Emformer removes the duplicated computation from the left context block by caching the key and value in previous segments' self-attention. Second, rather than passing the memory bank within the current layer in AM-TRF, inspired by transformer-xl~\cite{dai_2019} and its applicatin in speech recognition~\cite{liang2020interspeech}, Emformer carries over the memory bank from the lower layer. Third,  Emformer disables the summary vector's attention with memory bank to avoid overweighting the most left part of context information. Finally, Emformer applies a parallelized block processing training method, which is important to train Emformer for low latency speech recognition. 

To verify the performance of the proposed method, we carry out experiments on LibriSpeech~\cite{Panayotov2015}. More experiments using industry dataset with variant scenarios are in~\cite{yongqiang2021_icassp}. Under the average latency of 640 ms constraint, comparing with AM-TRF, Emformer gets relative WER reduction $17\%$ on test-clean and $9\%$ on test-other. Meanwhile, Emformer reduces the training time by almost $80\%$ and decoding RTF by $18\%$. For a low latency scenario with an average latency of 80 ms, Emformer saves more than $91\%$ computation from AM-TRF and obtain WER $3.01\%$ on test-clean and $7.09\%$ on test-other. According to our knowledge, this is the first work to give streaming transformer results on LibriSpeech with such low latency. Under the average latency of 960 ms and 640 ms constraint, Emformer also gives the best result on LibriSpeech so far.

\vspace{-7pt}
\section{Emformer}
\vspace{-5pt}
Emformer improves over the AM-TRF. The following subsection gives a short introduction to AM-TRF.
\vspace{-8pt}
\subsection{AM-TRF}
\vspace{-5pt}
Figure~(\ref{fig:amtrf}) illustrates the operations in one AM-TRF layer. A sequence of input feature vectors are chunked into multiple non-overlapping segments  $\mathbf{C}_1^n, \cdots, \mathbf{C}_{I-1}^n$, where the $i$ denotes the index of segment, and  $n$  the layer's index. In order to reduce boundary effect, left and right contextual blocks, $\mathbf {L}_i^n$ and $\mathbf{R}_i^n$, are concatenated with $\mathbf C_i^n$ to form a contextual segment $\mathbf{X}_i^n = [\mathbf {L}_i^n, \mathbf {C}_i^n, \mathbf R_i^n]$.
At the $i$-th segment, the $n$-th AM-TRF layer accepts $\mathbf X_i^n$ and a bank of memory vector $\mathbf{M}_i^n = [\mathbf{m}_1^n, \cdots, \mathbf{m}_{i-1}^n]$ as the input, and produces $\mathbf X_{i}^{n+1} = [\mathbf L_{i}^{n+1},\mathbf C_{i}^{n+1}, \mathbf R_{i}^{n+1}]$ and $\mathbf {m}_i^n$ as the output, whereas $\mathbf X_{i}^{n+1}$ is feed to the next layer and $\boldsymbol m_i^n$ is inserted into the memory bank to generate $\mathbf M_{i+1}^n$ and carried over to the next segment. After all the AM-TRF layers, the center blocks $\{\mathbf {C}_i^{N-1}\}_{i=0}^{I-1}$ are concatenated as the encoder output sequence; the contextual blocks $\{\mathbf L_i^{N-1}\}_{i=0}^{I-1}$ and $\{\mathbf R_i^{N-1}\}_{i=0}^{I-1}$ are discarded.

At the core of each AM-TRF layer, there is a modified attention mechanism which attends to the memory bank and yields a new memory vector at each segment: 
\vspace{-5pt}
\begin{align}
    \hat{\mathbf X}_i^n =& \mathrm{LayerNorm}(\mathbf X_i^n) \\
      \mathbf{K}_i^n=&\mathbf{W}_{\rm k}[\mathbf{M}_i^n,\hat{\mathbf X}_i^n], \label{amtrf-k} \\
      \mathbf{V}_i^n=&\mathbf{W}_{\rm v}[ \mathbf{M}_i^n,\hat{\mathbf X}_i^n], \label{amtrf-v} \\
    [\mathbf Z_{\mathrm{L}, i}^n, \mathbf Z_{\mathrm{C}, i}^n, \mathbf Z_{\mathrm{R}, i}^n] =& \mathrm{Attn}(\mathbf{W}_{\rm q}\hat{\mathbf X}_i^n, \mathbf{K}_i^n, \mathbf{V}_i^n) + \mathbf X_i^n \\
    \mathbf m_i^n =& \mathrm{Attn}(\mathbf{W}_{\rm q}\mathbf{s}_i^n, \mathbf{K}_i^n, \mathbf{V}_i^n) \label{eqn:mem-amtrf}
\end{align}
whereas $\mathbf Z_{\mathrm{L}, i}^n, \mathbf Z_{\mathrm{C}, i}^n$ and $\mathbf Z_{\mathrm{R}, i}^n$ are the attention output for $\mathbf L_i^n, \mathbf C_i^n$ and  $\mathbf R_i^n$ respectively; $\mathbf{s}_i^n$ is the mean of center block $\mathbf{C}_i^n$;  $\mathrm{Attn}(\mathbf q; \mathbf k, \mathbf v)$ is the attention operation defined in~\cite{vaswani_2017} with $\mathbf q$ , $\mathbf k$ and  $\mathbf v$ being the query, key and value, respectively. 

$\mathbf Z_{\mathrm{L}, i}^n, \mathbf Z_{\mathrm{C}, i}^n, \mathbf Z_{\mathrm{R}, i}^n$ are passed to a point-wise feed-forward network (FFN) with layer normalization and residual connection to generate the output of this AM-TRF layer, i.e., 
\begin{align}
\hat{\mathbf X}_{i+1}^{n} = \mathrm{FFN}(\mathrm{LayerNorm}([\mathbf Z_{\mathrm{L}, i}^n, \mathbf Z_{\mathrm{C}, i}^n, \mathbf Z_{\mathrm{R}, i}^n])) \\
\mathbf X_i^{n+1} = \mathrm{LayerNorm}(\hat{\mathbf X}_{i}^{n+1} + [\mathbf Z_{\mathrm{L}, i}^n, \mathbf Z_{\mathrm{C}, i}^n, \mathbf Z_{\mathrm{R}, i}^n]) \label{eqn:last_ln}
\end{align}
where FNN is a two-layer feed-forward network with Relu non-linearity.
The last layer normalization in Eq.~(\ref{eqn:last_ln}) is used to prevent a path to bypass all the AM-TRF layers. 

\comment{
\begin{figure}
    \centering
    \includegraphics[scale=0.4]{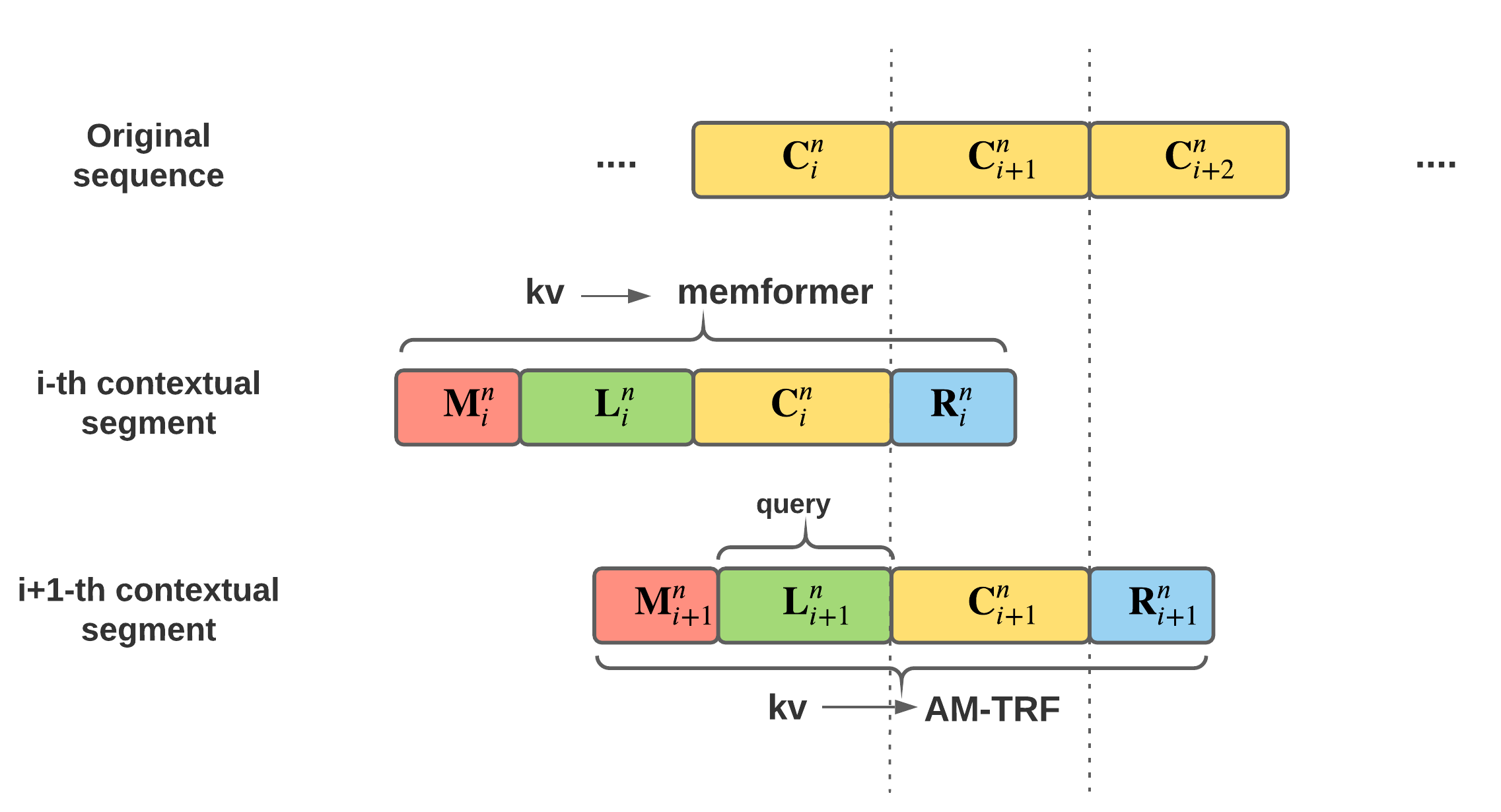}
    \vspace{-2em}
    \caption{Redundant computation in AM-TRF: for the same query $\mathbf L^n_{i+1}$, attention is performed over the new contextual segment thus need to recompute the attention output, while in Emformer, attention is performed over the old segment, thus can be cached. }
    \label{fig:limit}
\end{figure}
}

\vspace{-7pt}
\subsection{Emformer}
\vspace{-5pt}
As shown in~\cite{Chunyang_2020_interspeech}, given the similar latency constraint, AM-TRF has outperformed previous streaming transformer models. However, there are several issues with AM-TRF. The usage of the left context is not efficient. AM-TRF training relies on the sequential block processing that is not suitable for low latency model training. Having observed these limitations, we proposed a new streamable transformer architecture, namely, Emformer. One layer of {Emformer} is demonstrated in Figure~(\ref{fig:memformer}). The following subsections describe the important improvements made in Emformer. 

\vspace{-7pt}
\subsubsection{Cache key and value from previous segments}
\vspace{-5pt}
As illustrated in Figure~(\ref{fig:amtrf}), for the $i$-th segment, the embedding of the left context $\mathbf L_{i}^n$ needs to be re-computed for every step, even though $\mathbf L_{i}^n$ is overlapped with $\mathbf C_{i-1}^n$ (or possibly even more previous center blocks). Thus we only need to cache the projections from the previous segments. As shown in Figure~(\ref{fig:memformer}), Emformer only computes the \emph{key}, \emph{value} projections for the memory bank, center, and right context; Emformer saves the computation of \emph{query} projection of left context, as it does not need to give output from the left context block for the next layer. Compared with AM-TRF, the attention part in Emformer operates in the following sequence:
\begin{align}
 [\hat{\mathbf C}_i^n, \hat{\mathbf R}_i^n] &= \mathrm{LayerNorm}([\mathbf C_i^n, \mathbf R_i^n]) \\
\mathbf{K}_i^n&=[\mathbf{W}_{\rm k}\mathbf{M}_i^n, \mathbf{K}_{L,i}^n, \mathbf{W}_{\rm k}\mathbf{C}_i^n, \mathbf{W}_{\rm k}\mathbf{R}_i^n], \\
\mathbf{V}_i^n&=[ \mathbf{W}_{\rm v}\mathbf{M}_i^n, \mathbf{V}_{L,i}^n, \mathbf{W}_{\rm v}\mathbf{C}_i^n, \mathbf{W}_{\rm v}\mathbf{R}_i^n], \\
\mathbf Z_{\mathrm{C}, i}^n &= \mathrm{Attn}(\mathbf{W}_{\rm q}\hat{\mathbf C}_i^n, \mathbf{K}_i^n, \mathbf{V}_i^n) + \mathbf C_i^n \\
\mathbf Z_{\mathrm{R}, i}^n &= \mathrm{Attn}(\mathbf{W}_{\rm q}\hat{\mathbf R}_i^n, \mathbf{K}_i^n, \mathbf{V}_i^n) + \mathbf R_i^n \\
\mathbf m_i^n &= \mathrm{Attn}(\mathbf{W}_{\rm q}\mathbf s_i^n; \mathbf{K}_i^n, \mathbf{V}_i^n)
\end{align}
where $\mathbf{K}_{L,i}^n$ and $\mathbf{V}_{L,i}^n$ are the \emph{key} and \emph{value} copies from previous segments with no additional computations.

Let's assume \rm{L}, \rm{C}, \rm{R}, and \rm{M} are the lengths for the left context block, the center context, the right context, and the memory bank. 
the number of heads in the multi-head self-attention is \rm{h} and per head dimension is \rm{d}. Note the summary vector is the mean of the center segment, of which length is always 1. In practice, the memory bank is implemented in ring buffer way with small length, and the model dimension, \rm{dh}, is much larger than any of \rm{L, C, R}, and \rm{M}. Emformer saves approximately $\frac{\rm L}{\rm L + C +R}$ of AM-TRF computation. For low latency scenario with center context length 80 ms, right context length 40 ms, and left context length 1280 ms, Emformer reduces more than $91\%$ computation from AM-TRF. 
\comment{
\begin{table}[tbh]
    \centering
    \caption{Comparison of computational complexity of AM-TRF versus Emformer.}
    \begin{tabular}{|c|cc|}
    \hline
    Model      & Ops & Computation  \\
    \hline\hline
    \multirow{4}{*}{AM-TRF} & QKV-Proj. & (2M + 3L + 3C + 3R + 1)(dh)^2\\
                            & Att.      & (M+L+C+R+1)(L+C+R+1)(dh) \\
                            & Out-Proj. & (L+C+R+1)(dh)^2 \\
                            & FFN       & 8(L+C+R+1)(dh)^2 \\ 
    \hline
    \multirow{4}{*}{Emformer} & QKV-Proj. & (2M + 3C + 3R + 1)(dh)^2\\
                            & Att.      & (M+L+C+R+1)(L+C+R+1)(dh) \\
                            & Out-Proj. & (C+R+1)(dh)^2 \\
                            & FFN       & 8(C+R)(dh)^2 \\
    \hline
    Ratio & -- & $\sim\frac{\mathrm {C+R}}{\mathrm {L+C+R}}$ \\
    \hline
     \end{tabular}
    \label{tab:comp_factor}
\end{table}
}


\vspace{-7pt}
\subsubsection{Carryover memory vector from previous segments in the lower layer}
\vspace{-5pt}
The attention output from the summary vector $\mathbf s_i^n$ is a memory vector in the memory bank. The memory bank carries all the previous context information for future segments. As we can see from Figure~(\ref{fig:amtrf}), the memory vector $\mathbf m_i^n$ from the $i$-th segment in the $n$-th layer is a prerequisite for the $(i+1)$-th segment from the same layer. In training, the auto-regression characteristic of AM-TRF forces the block processing to be in a sequential way that is not suitable for GPU computing. Especially for low latency model training, where the center segment is small, sequential block processing chunks the whole utterance computation into a small computation loop, which renders extremely low GPU usage.

To support parallelization for block processing training, Emformer takes the memory bank input from previous segments in the lower layer rather than the same layer. In this way, for each Emformer layer, the whole sequence is trained in parallel, fully taking advantage of the GPU computing resources.

\vspace{-7pt}
\subsubsection{Disallow attention between the summary vector with the memory bank}
\vspace{-5pt}
According to Eq.~(\ref{eqn:mem-amtrf}), the memory vector is a weighted interpolation of \emph{values} projected from the memory bank, the left context block, the center block, and the right context block. For both AM-TRF and Emformer, assigning the attention weight between the summary vector and the memory bank to zero stabilizes the training and improves recognition accuracy for long-form speech. Including the memory bank information in the current memory vector cause the most left context information over-weighted. Similar to a recurrent neural network, enable the connection of summary vector with the memory back could cause gradient vanishing or explosion. For AM-TRF, the usage of the weak-attention suppression  method~\cite{Chunyang_2020_interspeech,yangyangshi_interspeech_2020} partially addresses the problem by setting weak-attention weights to zero.

\vspace{-7pt}
\subsubsection{Deal with look-ahead context leaking}
\vspace{-5pt}
The sequential block processing in AM-TRF training chunks the input sequence physically. The right context size bounds the look-ahead reception field. However, sequentially processing blocks significantly slows the training.
\begin{figure}[h!]
\vspace{-12pt}
   \begin{center}
    \includegraphics[width=2.5in]{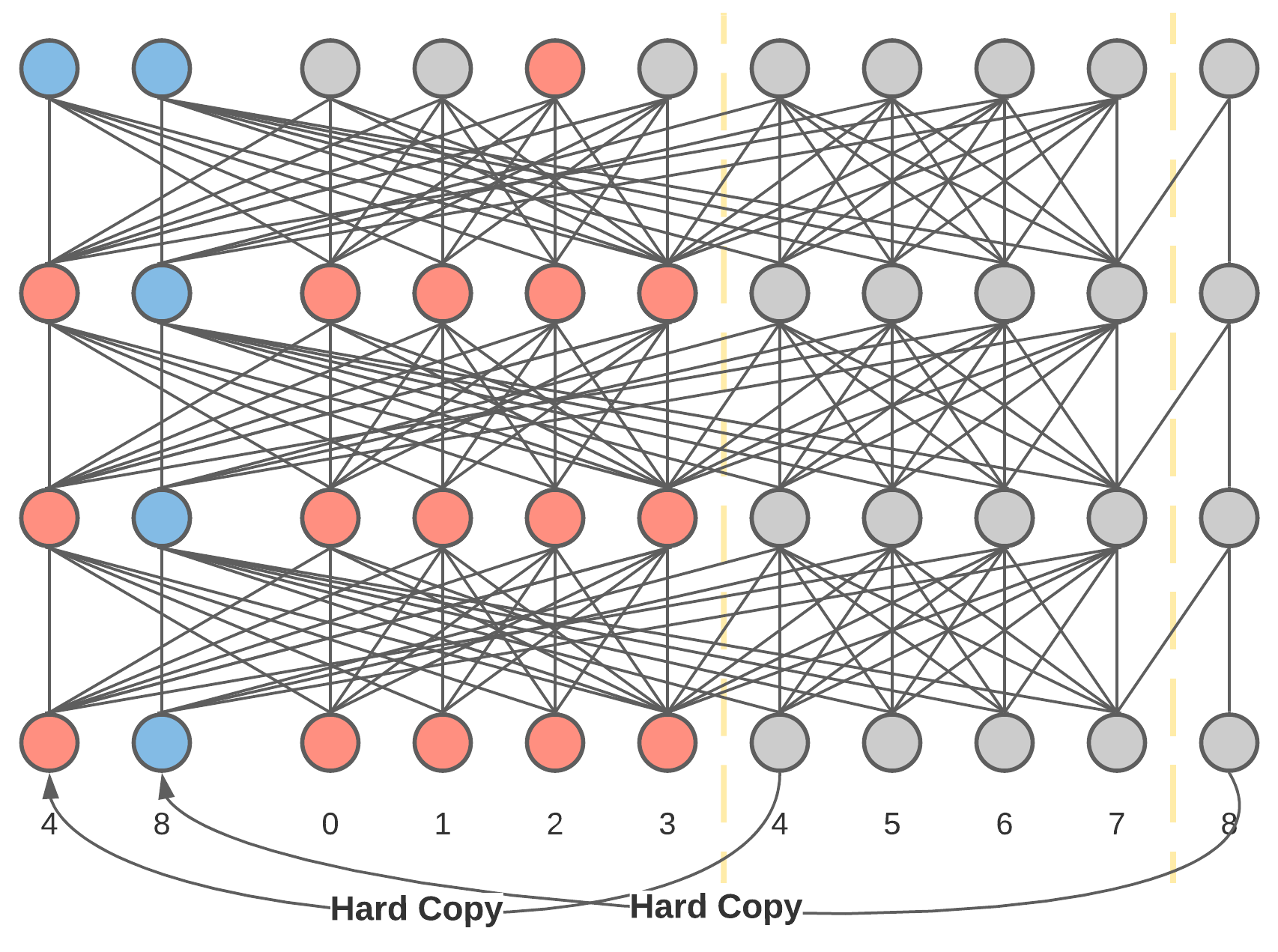}
    \vspace{-8pt}
    \end{center}
    \caption{Illustration of avoiding look-ahead context leaking. The chunk size is 4. The right context size is 1. }
    \label{fig:right_context_leaking}
    \vspace{-8pt}
\end{figure}
Now Emformer processes the input sequence in a fully parallel manner in the training stage. Like ~\cite{dai_2019,xiechen_2020_icassp}, Emformer applies attention masks to limit the reception field in each layer without physically chunking the input sequence. However, this method has the risk of a look-ahead of context leaking. The essential right context size grows when multiple transformer layers stack on top of one another. To deal with the look-ahead context leaking issue in training, Emformer makes a hard copy of each segment's look-ahead context and puts the look-ahead context copy at the input sequence's beginning as illustrated in the right part of Figure~\ref{fig:right_context_leaking}. For example, the output at the frame 2 in the first chunk only use the information from the current chunk together with the right context frame 4 without right context leaking.

\vspace{-7pt}
\section{Experiments}
\vspace{-8pt}
\subsection{Data and Setup}
\vspace{-5pt}
We verify the proposed method on the LibriSpeech corpus~\cite{Panayotov2015}. LibriSpeech has 1000 hours of book reading utterances derived from the LibriVox project. There are two subsets of development data and evaluation data in LibriSpeech.  The ``clean" subsets contain simple and clean utterances. The ``other" subset contains complex and noisy utterances. Based on the WER on the dev data, we select the best model and report its WER on test data. In the experiments, Emformer is used as an encoder for both the hybrid~\cite{yongqiang_2019_icassp,Chunyang_2020_interspeech,frank_zhang_2020} and transducer~\cite{zhang_2020,Yeh_2019,anmol_gulati_2020} models. 

\vspace{-7pt}
\subsubsection{Hybrid model} 
\vspace{-5pt}
The context and positional dependent graphemes are used as output units~\cite{le2019senones}. We use the standard Kaldi~\cite{Povey2011} LibriSpeech recipe to build bootstrap the HMM-GMM system. The 80-dimensional log Mel filter bank features at a 10 ms frame rate are used. We also apply speed perturbation~\cite{ko2015audio} and \emph{SpecAugment}~\cite{park2019specaugment} without time warping to stabilize the training. 

A linear layer maps the 80-dimensional features to 128 dimension vectors. Four continuous 128-dimensional vectors are concatenated with stride 4 to form a 512 vector that is the input to Emformer. In Emformer, each layer has eight heads of self-attention. The input and output for each layer have 512 nodes. The inner-layer of FFN has dimensionality 2048. Dropout is 0.1 for all layers across all experiments. For medium latency, memory bank length is 4. For low latency experiments where the segment size is small, memory bank information largely overlaps with direct left context. Therefore, we set the memory bank length to 0. An auxiliary incremental loss~\cite{Andros2019} with weight 0.3 is used to overcome the training divergence issue for deep transformer models. All hybrid models are trained with the adam optimizer~\cite{kingma2014adam} using 180 epochs. The learning rate increases to 1e-3 in 20K warming-up updates. Then it is fixed until 100 epochs. From then on, the learning rate shrinks every epoch with factor 0.95. All the models are trained using 32 Nvidia V100 GPUs with fp16 precision. We use hosts with Intel Xeon D-2191A 18-core CPUs to measure real time factors (RTFs). In measuring RTFs, 10 utterances are concurrently decoded. 

\vspace{-7pt}
\subsubsection{Transducer model}
\vspace{-5pt}
The output units are 1024 sentence pieces~\cite{kudo2018sentencepiece}  with byte pair encoding (BPE)~\cite{Sennrich_2016} as the segmentation algorithm. In the predictor, the tokens are first represented by 256-dimensional embeddings before going through two LSTM layers with 512 hidden nodes, followed by a linear projection to 640-dimensional features before the joiner.
For the joiner, the combined embeddings from the encoder and the predictor first go through a Tanh activation and then another linear projection to the target number of sentence pieces. Both the LCBLSTM and Emformer encoders are pre-trained from the hybrid systems. Similar to~\cite{anmol_gulati_2020}, we use a neural network language model (NNLM) for shallow fusion during beam search where the weight for NNLM probabilities was 0.3 across experiments. The training data for NNLM is the combined transcripts of the train set and the 800M text-only set.
\vspace{-7pt}
\subsection{Results}
\vspace{-5pt}
\subsubsection{Algorithmic latency induced by the encoder (EIL)}
\vspace{-5pt}
In block processing based decoding, the latency comes from the center block size and the look-ahead context size. For the most left frame in the center block, the latency is the center block size plus look-ahead context size. The latency for the most right frame in the center block is look-ahead context size. Therefore, we use algorithmic latency induced by the encoder (EIL), an average latency of all the frames in the center block, which equals to the look-ahead context latency plus center block latency discounted by 0.5.

\vspace{-7pt}
\subsubsection{From AM-TRF to Emformer}
\vspace{-5pt}
Table~\ref{tab:amtrf_to_mem} gives a performance comparison of AM-TRF with Emformer with a latency of 960 ms. Caching the key and value computation speeds up the training from 1.14 hours per epoch to 0.5 hours per epoch and decoding from RTF (real-time factor) 0.19 to 0.17. The left context caching also reduces the redundant gradient in training that results in some WER reduction\footnote{For large datasets, the caching strategy does not give WER reduction.}. Finally, using all improvements, comparing with AM-TRF, Emformer speeds up the training by 4.6 folds. Emformer also gets relative WER reduction $17\%$ on test-clean, $9\%$ on test-other and $18\%$ relative RTF reduction in decoding. For a low latency scenario, Emformer saves up to $91\%$ of computations from AM-TRF without considering parallel block processing. It is impractical to train AM-TRF for a low latency scenario. Therefore we ignore the detailed comparison.
\begin{table}[htb]
\vspace{-8pt}
    \centering
    \begin{tabular}{|l|c|cc|c|}
    \hline
    \multirow{2}{*}{Model}       & \multirow{2}{*}{RTF} & \multicolumn{2}{|c|} {test}  & train hours  \\
                &     & clean & other & per epoch \\
    \hline
    AM-TRF-24L      & 0.16 &3.27 & 6.66 & 1.14h \\
    \quad + left context caching & 0.13    &2.88 &6.44&0.50h  \\
    EM-24L &0.13 & 2.72 & 6.01 & 0.25h\\
     \hline
    \end{tabular}
    \caption{From AM-TRF to Emformer based on hybrid systems. All models have 80M parameters. Left context size, center block size and right context size are 640 ms, 1280 ms and 320 ms, respectively.}
    \label{tab:amtrf_to_mem}
    \vspace{-8pt}
\end{table}
\vspace{-10pt}
\subsubsection{Results from hybrid systems}
\vspace{-6pt}
\begin{table}[ht]
\vspace{-8pt}
    \centering
    \begin{tabular}{|l|cc|cc|c|}
    \hline
    \multirow{2}{*}{Model}   & \multirow{2}{*}{LC size} & \multirow{2}{*}{Center Size}  & \multicolumn{2}{|c|} {test} &  \multirow{2}{*}{RTF}    \\
                             &   &        & clean & other &  \\
    \hline\hline
    \multirow{2}{*}{LCBLSTM} & -- & 1280      &2.90 & 6.76 & 0.25 \\
                             & -- & 640       & 2.96 & 6.97 & 0.27 \\
    \hline
    \multirow{6}{*}{EM-24L} &  320  & \multirow{3}{*}{1280}    &2.75 & 6.08 & 0.13 \\
                             &  640  &    &2.72 & 6.01 & 0.13 \\
                             &  1280 &   &2.59 & 5.90 & 0.13 \\
                             \cline{2-6}
                             & 320  & \multirow{3}{*}{640} & 2.80 & 6.47 & 0.13 \\
                             & 640  &   & 2.78 & 6.46 & 0.13 \\
                             & 1280 &   & 2.76 & 6.59 &  0.15    \\
    \hline
    EM-36L &  \multirow{2}{*}{1280}   & \multirow{2}{*}{1280}    &2.58 & 5.75 & 0.17 \\
    \quad +smbr  &                     &                          &  \textbf{2.50} & \textbf{5.62} & 0.17  \\
    \hline
    EM-36L & \multirow{2}{*}{1280}   & \multirow{2}{*}{640}    & 2.69 & 6.14 & 0.20 \\
    \quad +smbr  &                     &                          &  2.62 & 5.97 & 0.19  \\
    \hline
    \end{tabular}
    \caption{Impact of left context (LC) size (in millisecond) on WER and RTF under medium latency constraints for hybrid models. Look-ahead size is 320 ms, the EIL is 640 ms or 960 ms when center size is 640 ms and 1280 ms, respectively. Both LCBLSTM and EM-24L have the similar 80M parameters. EM-36L has 120M parameters.}
    \label{tab:medium_latency}
    \vspace{-5pt}
\end{table}
Table~\ref{tab:medium_latency} and Table~\ref{tab:small_latency} presents the performance of the Emformer based hybrid systems for medium latency and low latency, respectively. For both tables, larger left context size gives better WER and slightly worse decoding RTF. In Table~\ref{tab:medium_latency}, LCBLSTM consists of 5 layers with 800 nodes in each layer each direction. Using a similar model size and latency constraint, Emformer gets a relative $48\%$ RTF deduction. Under EIL 1280 ms, Emformer obtained over relative $12\%$ WER  reduction over LCBLSTM on both test-clean and test-other datasets. Together with sMBR training~\cite{vesely2013sequence}, the Emformer with 120M parameters achieves WER $2.50\%$ on test-clean and $5.62\%$ on test-other under EIL 960 ms, and $2.62\%$ on test-clean and $5.97\%$ on test-other under EIL 640 ms.

In Table~\ref{tab:small_latency}, the LSTM consists of 7 layers with 1200 nodes in each layer. The input to LSTM is a concatenation of the current frame with eight look-ahead context frames. Low latency speech recognition gives higher RTF than medium latency speech recognition. Because medium latency speech recognition chunks an utterance into fewer larger segments, it speeds up the neural network's computation.  Using a similar model size and latency constraint, Emformer gets relative WER reduction $9\%$ and $15\%$ on test-clean and test-other, respectively. Together with sMBR training~\cite{vesely2013sequence}, the 36 layer Emformer achieves WER $3.01\%$ on test-clean and $7.09\%$ on test-other. According to our knowledge, for low latency 80 ms, Emformer gives the best WER on LibriSpeech data.
\begin{table}[htb]
\vspace{-8pt}
    \centering
    \begin{tabular}{|l|cc|cc|c|}
    \hline
    \multirow{2}{*}{Model}   & {LC size} & {latency}  & \multicolumn{2}{|c|} {test} &  \multirow{2}{*}{RTF}    \\

                        &     \multicolumn{2}{c|}{in milliseconds}           & clean & other &  \\
    \hline\hline
    LSTM & --     & 80  &3.75 & 9.18 & 0.25 \\
    \hline
    \multirow{3}{*}{EM-24L} &  320  & \multirow{3}{*}{80}   & 3.44 & 8.37 & 0.30 \\
                             &  640   &    & 3.37 & 8.05 & 0.31 \\
                             &  1280  &    & 3.41 & 7.75 & 0.33 \\
    \hline
    EM-36L &  \multirow{2}{*}{1280} & \multirow{2}{*}{80}  & 3.32 & 7.56 & 0.49 \\
    \quad +smbr  &                  &  &  \textbf{3.01} & \textbf{7.09} & 0.49 \\
    \hline
    \end{tabular}
    \caption{Impact of left context (LC) size (in millisecond) on word error rate and RTF under a low latency constraint for hybrid models. The look-ahead and center context size are 40 ms and 80 ms, respectively. Latency is defined by encoder induced latency (EIL). Both LSTM and EM-24L have the similar 80M parameters. EM-36L has 120M parameters. }
    \label{tab:small_latency}
    \vspace{-10pt}
\end{table}
\vspace{-10pt}
\subsubsection{Results from transducer systems}
\vspace{-5pt}
Table \ref{tab:transducer} summarizes the comparison between LCBLSTM and Emformer as encoders in the transducer system. 
Similar to the previous observations with hybrid systems, we see that given the same EIL (640 ms), Emformer consistently outperforms LCBLSTM on WER.
With the external NNLM, the transducer systems achieved similar WER to those from hybrid systems.

\begin{table}[htb]
\vspace{-5pt}
    \centering
    \begin{tabular}{|c|c|cc|}
    \hline
    Model     & NNLM  & \multicolumn{2}{|c|} {test}   \\
              &        & clean & other   \\
    \hline
    \multirow{2}{*}{LCBLSTM}   & \xmark     & 3.04 & 8.25  \\
          & \cmark    & 2.65 & 7.26  \\
    \hline
    \multirow{2}{*}{EM-24L}   & \xmark    & 2.78 & 6.92  \\
           & \cmark    & 2.37 & 6.07  \\
    \hline
    \end{tabular}
    \caption{WER of Emformer with the neural transducers. Both models use an EIL 640 ms with center context 640 ms and look-ahead context 320 ms. Left context size is 1280 ms.
    }
    \label{tab:transducer}
    \vspace{-10pt}
\end{table}
\vspace{-10pt}
\section{Conclusions}
\vspace{-5pt}
The proposed Emformer applied a cache strategy to remove the duplicated computation in augmented memory transformer (AM-TRF) for the left context. Emformer disabled the summary vector attention with a memory bank to stabilize the training. By redefining the memory carryover procedure and avoiding the right context leaking, Emformer supported parallelized block processing in training. Comparing with AM-TRF, Emformer got $4.6$ folds of training speedup and $18\%$ decoding RTF reduction. Experiments on LibriSpeech showed that Emformer outperformed the baselines in both hybrid and transducer systems. Under average latency EIL 960 ms, Emformer achieved WER $2.50\%$ on test-clean and $5.62\%$ on test-other with decoding RTF 0.13. Under low latency 80 ms constraint, Emformer achieved WER $3.01\%$ on test-clean and $7.09\%$ on test-other.
\vfill\pagebreak

\bibliographystyle{IEEEbib}
\bibliography{icassp_2021}

\end{document}